\documentclass{aa}  

\usepackage{natbib}
\usepackage{graphicx}
\usepackage{txfonts}
\usepackage{color}
\definecolor{red}{rgb}{1,0,0}
\definecolor{blue}{rgb}{0,0,1}

\usepackage{amstext}

\begin{document}

   \title{The inner dust shell of \object{Betelgeuse} detected by polarimetric aperture-masking interferometry \thanks{based on SAMPol data obtained at the ESO VLT Yepun telescope (090.D-0898(A)).}
   %A new look at \object{Betelgeuse} in the $H$ band % convection cells..spotty convection face.?
   }

\titlerunning{Inner dust shell of Betelgeuse.}

%   \subtitle{I. Overviewing the $\kappa$-mechanism}

   \author{X. Haubois\inst{1,2}
         \and
         B. Norris\inst{2}
          \and
              P.G. Tuthill\inst{2}
                \and         
                  C. Pinte\inst{3,4}
        \and
       P. Kervella\inst{4,5}
                \and          
           J. H. Girard\inst{1}
                \and          
        N.M. Kostogryz \inst{6}
        \and 
        S.V. Berdyugina \inst{6}
        \and
        G. Perrin\inst{5}
          \and
          S. Lacour\inst{5}
              \and
                A. Chiavassa\inst{7}
\and
          S.T. Ridgway\inst{8}
          }

   \institute{$^{1}$ European Organisation for Astronomical Research in the Southern Hemisphere, Casilla 19001, Santiago 19, Chile
   \\    \email{xhaubois@eso.org} 
       \\ $^{2}$ Sydney Institute for Astronomy, School of Physics, University of Sydney, NSW 2006, Australia 
\\$^{3}$ UJF-Grenoble 1 / CNRS-INSU, Institut de Plan\'{e}tologie et d'Astrophysique de Grenoble, UMR 5274, 38041 Grenoble, France  
\\$^{4}$ Unidad Mixta Internacional Franco-Chilena de Astronom\'{i}a, CNRS/INSU UMI 3386 and Departamento de Astronom\'{i}a,
Universidad de Chile, Casilla 36-D, Santiago, Chile
\\$^{5}$  LESIA, Observatoire de Paris, PSL Research University, CNRS, Sorbonne Universit{\'e}s, UPMC Univ. Paris 06, Univ. Paris Diderot,
Sorbonne Paris Cit{\'e}, 5 place Jules Janssen, 92195 Meudon, France
\\$^{6}$  Kiepenheuer-Institut f\"{u}r Sonnenphysik (KIS), Sch\"{o}neckstrasse 6, 79104 Freiburg, Germany
 \\$^{7}$  Laboratoire Lagrange, Universit{\'e} C\^ote d'Azur, Observatoire de la C\^ote d'Azur, CNRS, Boulevard de l'Observatoire, CS 34229, 06304 Nice Cedex 4, France        
 \\$^{8}$ National Optical Astronomy Observatory, P.O. Box 26732, Tucson, AZ 85726-6732, USA  
} 
   \date{Received / Accepted }

% \abstract{}{}{}{}{}
% 5 {} token are mandatory
 
  \abstract
  {Theory surrounding the origin of the dust-laden winds from evolved stars remains mired in controversy. Characterizing the formation loci and the dust distribution within approximately the first stellar radius above the surface is crucial for understanding the physics that underlie the mass-loss phenomenon.}
   {By exploiting interferometric polarimetry, we derive the fundamental parameters that govern the dust structure at the wind base of a red supergiant.}
   {We present near-infrared aperture-masking observations of \object{Betelgeuse} in polarimetric mode obtained with the NACO/SAMPol instrument.  We used both parametric models and radiative transfer simulations to predict polarimetric differential visibility data and compared them to SPHERE/ZIMPOL measurements.}
   {Using a thin dust shell model, we report the discovery of a dust halo that is located at only 0.5 R$_{\star}$ above the photosphere (i.e. an inner radius of the dust halo of 1.5 R$_{\star}$). By fitting the data under the assumption of Mie scattering, we estimate the grain size and density for various dust species. By extrapolating to the visible wavelengths using radiative transfer simulations, we compare our model with SPHERE/ZIMPOL data and find that models based on dust mixtures that are dominated by forsterite are most favored. Such a close dusty atmosphere has profound implications for the dust formation mechanisms around red supergiants.}
   {}

\maketitle
%
%________________________________________________________________

\section{Introduction}
 \object{Betelgeuse} ($\alpha$ Orionis) is a red supergiant (hereafter RSG) of spectral type M2Iab and one of the brightest stars in the night sky at all wavelengths. With a distance of $222_{-34}^{+48}$ pc \citep{2017AJ....154...11H}, its angular diameter of $\sim$43 mas in the near-infrared  \citep[e.g., ][]{2004A&A...418..675P,2009A&A...508..923H,2014A&A...572A..17M} makes it an ideal target for studying the inner structures that are involved in the poorly constrained mass-loss phenomenon.
For an overview of Betelgeuse's properties, we refer to the proceedings of a workshop that focused on Betelgeuse \citep{2013EAS....60.....K}.

Betelgeuse exhibits a complex signal that is spectrally and temporally variable in net polarized light.  From the temporal analysis of four years of $B$-band polarization data, the reported variability has been interpreted in terms of activity of large-scale convective cells \citep{1984ApJS...55..179H}. UBV polarimetric measurements from 6'' to 25'' around the star \citep{1986A&A...168..211L} led to the detection of a silicon dust environment whose grain size is in the range between 0.05 and 0.5 microns \citep{1986A&A...168..217M}.

Interferometric reconstructed images in the visible unveiled an elliptical structure around Betelgeuse at $\sim$ 2-2.5 stellar radii \citep{1985ApJ...295L..21R}. The favored explanation for this circumstellar emission was Mie scattering of stellar light by dust particles. % rather than electron-scattering, since silicate dust is present in the IR spectra \citep[e.g. recently an excess located between 8 and 200 $\mu$m is reported in][]{2013EAS....60..191K}. 
This interpretation was found in agreement with \cite{1981ASSL...88..317D}, who showed that assuming a 3600K effective temperature for Betelgeuse, clean silicate grains would start to condense at 1.8 stellar radius. This confirmed previous work predicting that at such a short distance above the photosphere, this dust shell would yield a significant degree of polarized light in the visible \citep{1978PASJ...30..435T}. 
%Adding these previously cited observations with optical speckle imaging, \cite{1986ApJ...308..260K} claims a detection of two companions for Betelgeuse and re-interprets polarization variability as the signature of multiplicity.
Based on interferometric and spectroscopic observations, \cite{Verhoelst2006} suggested that an amorphous alumina shell at 1.5 $R_{\star}$ could account for the fact that Betelgeuse appears 1.5 times larger in the mid-infrared (MIR) than in the near-infrared (NIR).
However, given the sublimation temperature of 1900 K for alumina, the gas pressure required to maintain such a shell is $10^4$ times higher than model predictions for the inner atmosphere of Betelgeuse. \cite{2007A&A...474..599P} found that a layer of water vapor coexists with a thin shell composed of SiO and alumina in a maximum radius of about 62.5 mas, corresponding to 1.43 R$_{\star}$ in the $N$ band. Farther away at about arcsecond spatial scales, \cite{2011A&A...531A.117K} reported MIR spectroimaging observations of the highly asymmetric Betelgeuse circumstellar environment, concluding that silicates or alumina dust were viable constituents. 

The atmosphere of Betelgeuse is subject to $\sim$1G magnetic fields inferred from the Zeeman effect \citep{2010A&A...516L...2A}, which are thought to originate from local low-scale convective activity. 
Simplifying the interpretation of polarized-light signals, such modest fields are unlikely to result in significant levels of direct continuum polarization.
Extreme adaptive-optics polarimetric observations within the Spectro-Polarimetric High-contrast Exoplanet Research (SPHERE) obtained with the Z{\"u}rich Imaging Polarimeter (ZIMPOL) were obtained by \cite{2016A&A...585A..28K}, resolving the surface and inner environment of Betelgeuse.
A somewhat complex picture emerged from this work, in which media of inhomogeneous densities and temperatures coexist in asymmetric structures.
Dust is confirmed at 3 R$_{\star}$, but the authors note that polarization levels might be diluted by (unpolarized) gaseous emission, which allows for the possibility that significant dust may exist interior to this.

Developments in polarimetric aperture masking interferometry (SAMPol mode at NACO) allowed novel polarized-light studies of dust shells around AGB stars \citep{2012Natur.484..220N}. Observational studies targeting several bright systems revealed dust shells capable of strong scattered-light interaction with the stellar radiation field. These systems had typical diameters of about 2 $R_{\star}$ and grain sizes of about 300~nm. 
These data immediately provoked renewed interest in the role played by photon scattering in the physics of radiatively driven winds, opening a new window to explore links between the stellar properties and final mass-loss rates on the AGB.

Based on the same observing technique, we aim to identify the nature and the location of the polarizing structures at the wind base for the red supergiant Betelgeuse. We here report on optical interferometry polarimetric observations of Betelgeuse obtained with the NACO/SAMPol instrument. Sect.~\ref{obsred} describes the observations and the data reduction. In Sect.~\ref{envdet} we present the detection of a thin polarizing dust shell. In Sect.~\ref{rad} radiative transfer modeling is used to reproduce the SAMPol observations and compare them to the visible observations made with ZIMPOL. In Sect.~\ref{discuss} we discuss the effects of polarization that stem from the stellar atmosphere before we conclude in Sect.~\ref{conclusect}.

%\footnote{Observations obtained run ID 090.D-0898(A)}.
%Images of \object{Betelgeuse} reconstructed with the MIRA and SQUEEZE algorithms are compared before concluding in Sect.~\ref{conclu}.
%

\section{Observation and data reduction}
\label{obsred}

Betelgeuse was observed with NACO/SAMPol on 28 January 2013. 
The instrument was configured to perform aperture-masking interferometry (SAM mode) \citep{2003SPIE.4841..944L,2003SPIE.4839..140R} together with the use of a Wollaston prism and rotating half-wave plates (HWP), which together comprise the supported ``SAMPol mode''`.
Following the work of \cite{2012Natur.484..220N}, the extreme spatial resolution delivered by SAM in concert with differential polarimetric data is capable of revealing scattered-light structures at spatial scales that are not accessible to competing technologies.
Despite strongly polarized individual constituent components, an astrophysical image will often yield zero net polarization as a sum over a circularly symmetric environment. 
Operating at the extreme spatial resolution limit, SAMPol is capable of breaking the degeneracy, delivering information on polarized structures at spatial scales that are normally hidden.

This work employed the 18-hole aperture mask (net transmission of 3.9\% with respect to the full aperture), with an integration time of 0.1s obtained with a subframe window of 512$\times$514 pixels on the Aladdin3 detector. For each image data cube, interferograms for the two orthogonal polarizations split by the Wollaston prism are recorded simultaneously on the detector array. The full starlight polarization state is explored by changing the rotation angle of upstream half-wave plates. This experiment employed four half-wave plate positions separated by 22.5$\degr$ in order to sample the Q and U Stokes parameters. Following a relatively standard data reduction procedure entailing sky subtraction, flat fielding, and cosmic-ray removal, the complex visibilities and bispectra were separately accumulated for each baseline and baseline triplet, respectively, for each of the two polarizations. Finally, in order to remove systematics that are due to the non-common paths after the Wollaston prism, horizontally polarized visibilities (V$_{h}$) were divided by vertically polarized visibilities (V$_{v}$), forming a single differential observable. This ratio of the observables, hereafter called differential (polarimetric) visibility, is highly robust to errors induced by seeing because the two polarizations were recorded simultaneously. A list of spectral filters employed for the SAMPol observations is given in Table~\ref{tab:filters}.

% \rouge {ajouter NIRC observations si utile ? Observations spectro de NACO jHK avec longue fente en 2013, verifier si utile, elles sont dans les archives.}
\begin{table*}
\centering
\begin{tabular}{lccccl}
\hline
Filter & Central Wavelength ($\mu$m) & Width ($\mu$m)  & Observed Source(s)  \\
\hline
 NB 1.04 & 1.040 & 0.015 & Betelgeuse/Aldebaran \\ 
NB 1.08 & 1.083  & 0.015 & Betelgeuse \\ 
NB 1.09 & 1.094 & 0.015 & Betelgeuse/Aldebaran \\ 
NB 1.24 & 1.237 & 0.015 & Betelgeuse/Aldebaran \\ 
NB 1.28 & 1.282 & 0.014 & Betelgeuse \\ 
NB 1.64 & 1.644  & 0.018 & Betelgeuse/Aldebaran \\ 
NB 1.75 & 1.748 & 0.026 & Betelgeuse/Aldebaran \\ 
NB 2.12  & 2.122 & 0.022 & Betelgeuse/Aldebaran \\ 
IB 2.30 & 2.30 & 0.06  & Betelgeuse/Aldebaran \\ 
IB 2.36 & 2.36 & 0.06 & Betelgeuse/Aldebaran \\ 
IB 2.42 & 2.42 & 0.06 & Betelgeuse/Aldebaran\\
%NB 3.74 & 3.740  & 0.02 & Betelgeuse/Aldebaran\\
\hline
\end{tabular}
\caption{Characteristics of spectral filters employed for the SAMPol observations.}
\label{tab:filters}
\end{table*}

\section{Detection of the dusty inner environment of \object{Betelgeuse}}
\label{envdet}

\begin{figure}[h!]
\begin{center}
 \includegraphics[width=3.5in]{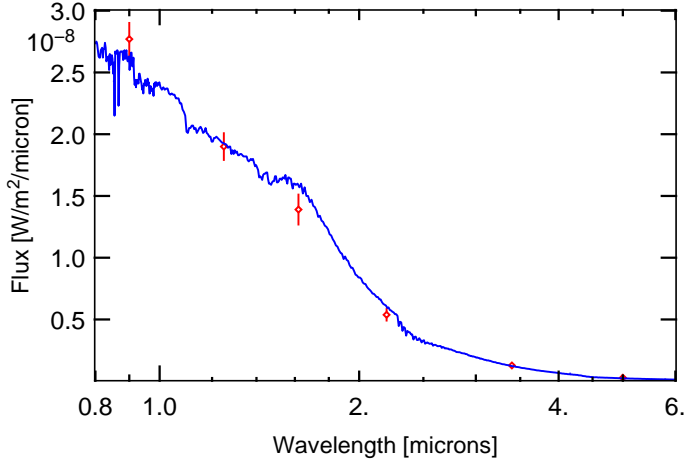}
\end{center}
\caption{Spectral energy distribution of Aldebaran. Red points correspond to observed magnitudes reported in \cite{2002yCat.2237....0D} The blue line shows a synthetic model with T$_{eff}$=3900 K, log g =1.5, and a limb-darkened disk diameter of 20.60 mas.}
\label{fig:Ald}
\end{figure}

Although strictly speaking, the self-calibrating differential polarimetric visibility obviates the need for the usual strict observance of known (usually point source) stars for the system transfer function, here the dust-free red giant Aldebaran was chosen as a non-polarized reference object.
Photometric measurements performed in the optical and infrared domains \citep{2002yCat.2237....0D} were compared with a simple (non-dusty) synthetic spectrum obtained from \cite{2003IAUS..210P.A20C}\footnote{Accessible from http://www.stsci.edu/hst/observatory/crds/} and with the following parameters: effective temperature T$_{eff}$=3900 K, log g =1.5 and limb-darkened disk with a diameter of 20.60 mas (Fig.~\ref{fig:Ald}). The $\chi^{2}$ per data point (equivalent of a reduced $\chi^{2}$) of the comparison between the Kurucz spectrum and the photometric measurements is equal to 1.95, meaning that no significant infrared excess is present in this part of Aldebaran's spectrum. Furthermore, no infrared excess was detected between 8 and 13 $\mu$m \citep{1998ApJ...502..833M}, nor between 50 and 670 $\mu$m \citep{2011A&A...533A.107D}.

\begin{figure*}
\centering
 \includegraphics[width=3.5in]{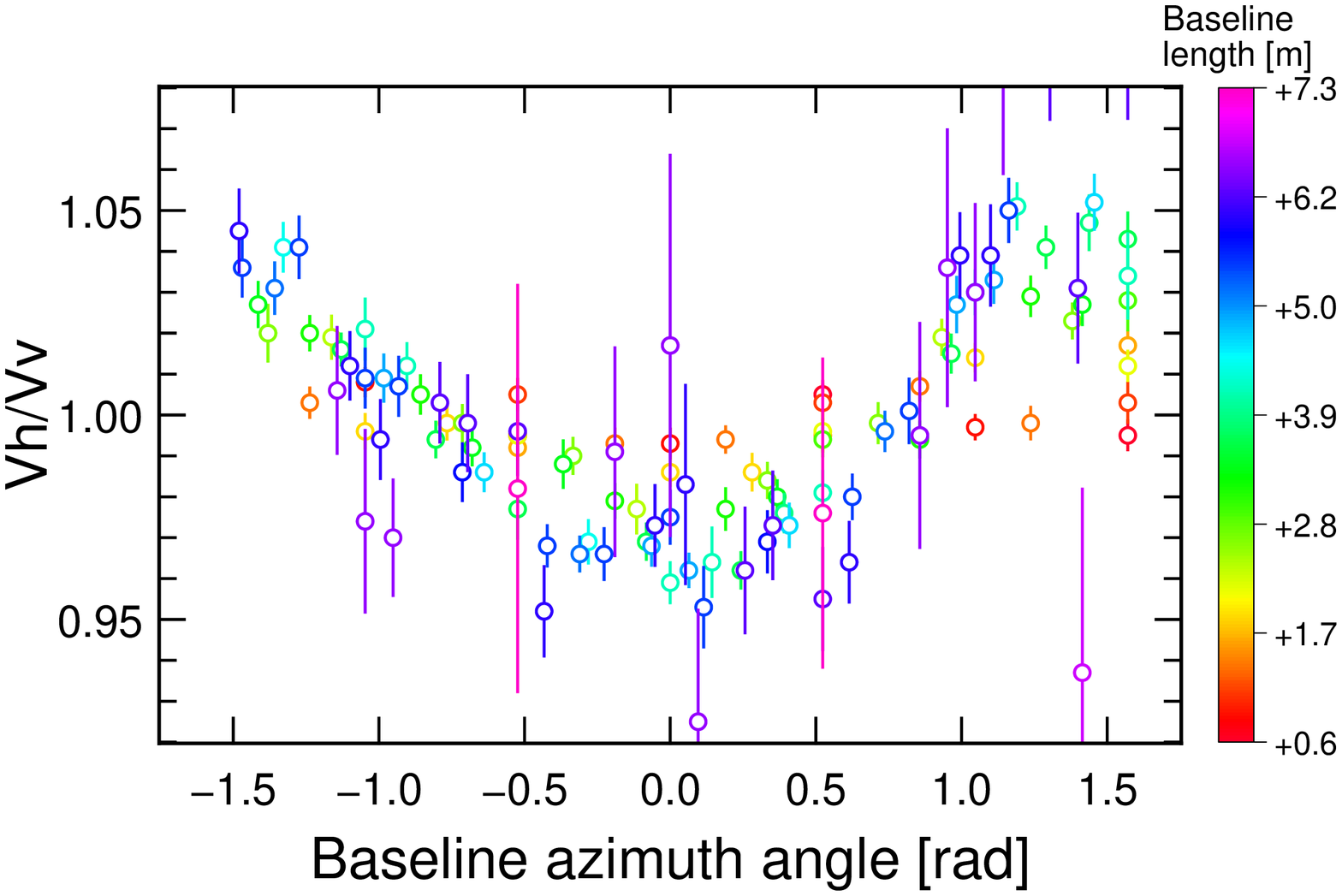}
 \hspace{3mm}
 \includegraphics[width=3.5in]{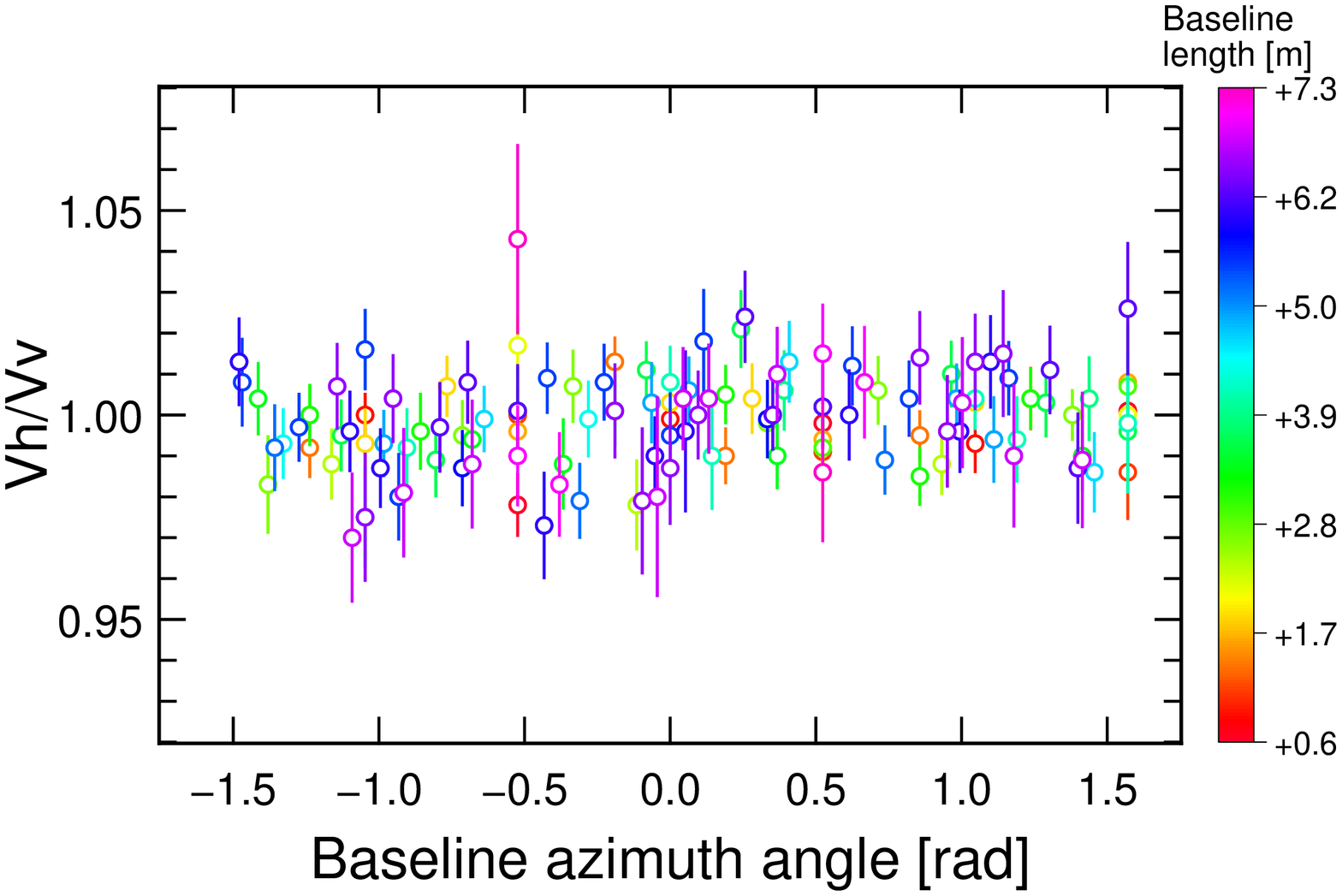}
\caption{Differential visibility ratio plotted as a function of azimuth angle, color-coded with baseline length for the filter centered on 1.04 $\mu$m. Left: Betelgeuse. Right: Aldebaran.}
\label{vhvv_1.04}
\end{figure*}

Differential polarized visibility (V$_{h}$/V$_{v}$) curves for Betelgeuse (left panel) and Aldebaran (right panel) are presented in Figure~\ref{vhvv_1.04} for the filter centered on 1.04 $\mu m$. The corresponding curves for other filters are presented in Figs. \ref{vhvv1} and \ref{vhvv2}. Any significant departure from the value of unity over a range of baselines indicates that a polarized structure has been resolved. The varying signal apparent for Betelgeuse (but not Aldebaran) can be first approximated by a sinusoid, and is the first-order expectation for any spherically symmetric structure \citep[see also][]{2012Natur.484..220N}.
We note that the amplitude of this sinusoidal variation decreases with observing wavelength: the filters with central wavelength above 1.75 $\mu$m show no sign of a polarized signal.
This is expected from signals with an astrophysical origin in which the dust grain-size distribution enforces a strong spectral dependence of the polarization. 
Although this provides a strong affirmation that the SAMPol data have reach into the physics of the circumstellar dust halo, it also means that the longer wavelength data with null signals have scant ability to constrain the models, and are therefore largely discarded in the subsequent analysis.

It is also worth emphasizing that compared to Betelgeuse, Aldebaran shows no significant variation of differential visibility, neither as a function of baseline length nor of azimuth angle.  The finding of a null result for polarimetric signals that are detected by observing a non-dusty source lends enormous confidence that no major instrumental polarization biases or uncalibrated systematic errors remain to contaminate the data when we consider our results on Betelgeuse.

%The curves for Aldebaran show a differential visibility ratio compatible with 1.
%SI Variations chez aldebaran , expliquer par CO clumps de Onhaka 2013?

\subsection{Dust shell radius}
\begin{figure}[h!!]
\begin{center}
 \includegraphics[width=3.in]{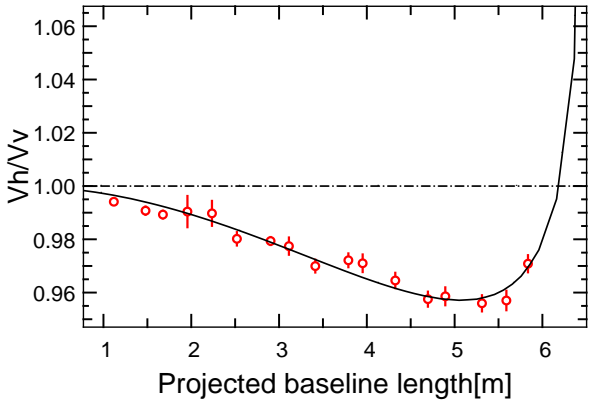}
\end{center}
\caption{Azimuthally reduced differential visibilities as a function of projected baseline length at 1.09 $\mu$m. SAMPol data are shown in red, and the dust shell model is plotted as a black line.}
\label{vhvv109}
\end{figure}

\begin{figure}[h!!]
\begin{center}
 \includegraphics[width=3.5in]{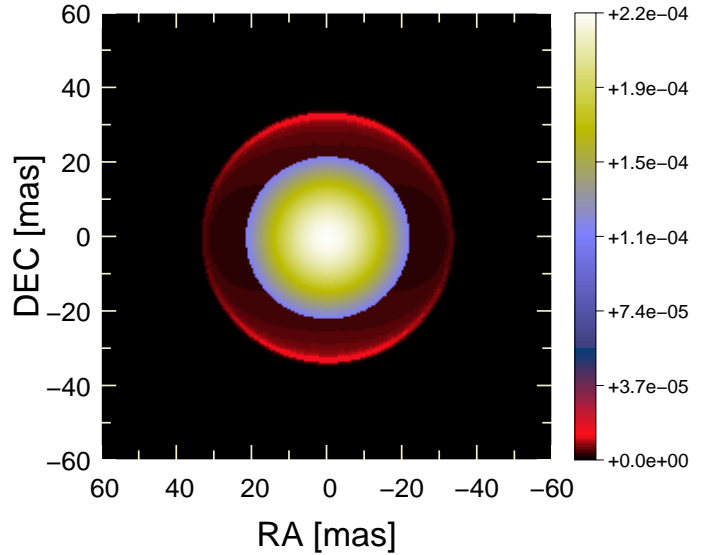}
\end{center}
\caption{Image of the dust shell model in the horizontal polarization at 1.09 $\mu$m. We used the parameters that correspond to the best fit as reported in Table~\ref{tab:dustshell}. The intensities are normalized to the total intensity of the image.}
\label{im109}
\end{figure}

The amplitude of the differential polarized visibility signal is about 4\% at 1.04 micron for Betelgeuse, in accord with theoretical predictions from NIR scattered-light models on a thin dust shell \cite[section 3.3 of][]{2009A&A...508..923H}.
Because the dominant term is a sinusoidal variation of the differential polarized visibility curves with azimuth (Figs.~\ref{vhvv_1.04}, \ref{vhvv1}, and \ref{vhvv2}), the simplest first-order model to produce such a polarized signal is a spherical dust shell that scatters the stellar light. This model entails only three parameters: the stellar uniform-disk radius $R_{\star}$, the dust shell radius $R_{dust}$ , and the flux ratio between the two structures. Hereafter we term the latter the ``scattered-light fraction'' because it represents a proxy for the ratio between the flux radiated from gas and the scattered flux from dust.
It is important to emphasize that we define this fraction for the polarized light alone. In this model, the shell thickness is negligible compared to its diameter.
 The model computes differential polarized visibilities from images that are simulated in the two orthogonal polarizations, allowing fits to the observed primary observables. This model was previously used in \cite{2005MNRAS.361..337I} and  in \cite{2012Natur.484..220N}.
Because simple circular symmetry is assumed, differential polarized visibilities could easily be reduced to a single azimuthally symmetric quantity, permitting the model to be parameterized as a function of baseline length alone.
An example best-fit model is presented in Fig.~\ref{vhvv109}, and the corresponding 2D image from this parametric model is shown in Fig.~\ref{im109}.
Model results for all filters are summarized in Table.~\ref{tab:dustshell}.

%Measurements were made only on Q stokes data, since U showed some high depolarization in many filters...

\begin{table*}
\centering
\begin{tabular}{lccccl}
\hline
Filter & $R_{\star}$ (mas) &   $R_{dust}$ (mas) & Scattered-light fraction [\%] & $\chi_{reduced}^{2}$ \\
\hline
NB 1.04 & 22.00  $\pm$ 0.24   & 32.35 $\pm$  0.17 & 4.4 $\pm$ 0.3 & 3.2 \\ 
NB 1.08 & 22.31   $\pm$ 0.23 & 34.53  $\pm$ 0.12 & 4.2 $\pm$ 0.1  & 0.5 \\ 
NB 1.09  & 22.31 $\pm$ 0.23  &  33.22 $\pm$ 0.10 & 4.1 $\pm$ 0.1  & 1.8 \\ 
NB 1.24 & 23.84 $\pm$ 0.34   & 29.20 $\pm$ 5.22 & 2.2 $\pm$ 0.8 & 4.9  \\ 
NB 1.64 & 22.36 $\pm$ 0.80   & 27.41 $\pm$ 4.02  & 2.0 $\pm$ 0.8 & 2.8 \\ 
NB 1.75 & 23.30  $\pm$ 0.90 &  29.62 $\pm$ 5.24 & 1.5 $\pm$ 0.5  & 2.2\\ 
\hline
\end{tabular}
\caption{Model parameters fit to the differential visibilities in the filters where a polarized signal was detected.}
\label{tab:dustshell}
\end{table*}

%Two weeks after our SAMPol observations, an $H$-band limb-darkened diameter estimation of 43.73$\pm$0.5 mas was obtained with PIONIER \citep[Fig~2. and Tab.~3 in ][]{2016A&A...588A.%130M}. Based on the position of the visibility-curve first null, this limb-darkened diameter (using a power-law index of 0.19 $\pm$ 0.07) corresponds to a 42.3$\pm$0.5 mas uniform-disk diameter, which does not agree with our 44.72$\pm$1.6 mas value (from Tab.~\ref{tab:dustshell}).

The model describing the underlying stellar brightness distribution was found to have very little impact on the fits.
Stellar models employing a uniform disk gave identical results to those assuming a limb-darkened disk. 
This can be explained by the relatively limited angular resolution intrinsic to the SAMPol data: no second and higher lobes of 
the visibility function are probed, where the detailed form of the brightness profile is important.

The uncertainty in the fit parameters increases with wavelength as the polarized signal decreases.
Taking into account the three short-wavelength spectral filters, we find that $R\_{dust} \approx 1.5~R_{\star} $.
The finding of strong scattered-light signals from dust in such proximity to the  stellar surface constrains the dust chemistry. In the following, we consider three types of dust species found in O-rich evolved stars, all of which have high sublimation temperatures: MgSiO$_{3}$ (enstatite), Mg$_{2}$SiO$_{4}$ (forsterite), and Al$_{2}$O$_{3}$ (alumina).

%David's condition for dust formation:
%Low temperatures (< 2000 K) to afford condensation
%Sufficient high densities (~1010 ? 1016 cm-3) to allow for coalescence
%Availability of atoms, molecules & clusters for the specific condensate
%Sufficient time for dust clusters and grains to grow (time scales)
%kobayashi, T_sublimation .

\subsection{Modeling the scattered-light fraction}
As previously presented in  \cite{2012Natur.484..220N}, the decrease in the scattered-light signal with increasing wavelength can be reproduced by a Mie scattering opacity.
Denoting photons scattered by dust as $I_{dust}$, we employ the same model, where the scattered-light fraction can be written as
\begin{equation}
\frac{I_{dust}}{I_{total}} = \frac{B(1-e^{-\tau_{sc}})}{e^{-\tau_{sc}}+B(1-e^{-\tau_{sc}})}  
,\end{equation}

where $\tau_{sc}$ is the product of the Mie scattering cross section (computed using numerical routines made available by the Earth Observation Data group from the University of Oxford\footnote{http://eodg.atm.ox.ac.uk/MIE/index.html}) and of the surface grain density.  The term $B$ takes into account the fraction of light that is intercepted by the star after it is back-scattered from the dust shell. The optical constants of the enstatite and forsterite species were derived from \cite{2003A&A...408..193J}  and taken from the Jena University Database\footnote{http://www.astro.uni-jena.de/Laboratory/OCDB/amsilicates.html}. For Al2O3, we used the optical constants distributed in the MCFOST \citep{2009A&A...498..967P} and ProDiMo \citep{2009A&A...501..383W} modeling codes.

\begin{figure}[h!!]
\begin{center}
 \includegraphics[width=3.2in]{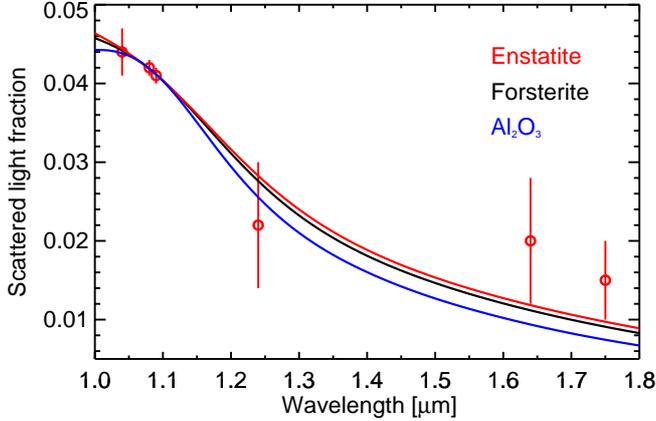}
\end{center}
\caption{Scattered-light fraction measured by SAMPol (red diamonds) with best-fit thin shell models based on the Mie scattering model for three dust species. The red error bars correspond to the 1~$\sigma$ uncertainty of the measurements.}
\label{scat_all_3}
\end{figure}

The outcome of this modeling, as shown in Figure~\ref{scat_all_3}, is that the SAMPol measurements are well reproduced by the thin dust shell model. Table~\ref{tab:Mie} presents the values of the best-fit parameters.
 
The size of the dust grains and the mass of the shell are similar to those for the AGB stars reported in \cite{2012Natur.484..220N}. For oxygen-rich AGB stars, one scenario proposes that photon scattering by Fe-free silicate grains can couple the momentum of the radiation field to the dust. This mechanism is able to launch a dusty wind provided the grain size is about 1 micron \citep{2008A&A...491L...1H}. Based on the grain size estimation derived from our data,  this scenario seems to be plausible for Betelgeuse as well and might even hold for other oxygen-rich red supergiants. However, the composition of such grains remains to be determined, and the question of how they can survive in such proximity to the stellar surface needs to be addressed.

  % Comments on the mass loss rate deduced from CO [alma ] ??
 % this dust shell might not explain it but at least might be fundamental for the mechanism that leads to the production of more dust. (How ? Sheltering from radiation is not plausible since it barely absorbs the stellar light )

\begin{table*}
\centering
\begin{tabular}{lcccccl}
\hline
Dust species &   grain radius (nm) &  Surface grain density ($10^{6}~cm^{-2}$) & Dust shell mass ($10^{-10}$ M$_\odot$)   & $\chi_{reduced}^{2}$ \\
\hline
Alumina & 285.7 $\pm$ 13.2 & 6.1  $\pm$ 1.1   &1.49 $\pm$ 0.13 &  1.1 \\ % D=222pc
Forsterite  &  320.0 $\pm$ 20.3  & 6.5 $\pm$ 1.6 & 1.58  $\pm$ 0.19 &  0.8  \\% D=222pc
Enstatite & 335.9 $\pm$ 24.5  &  6.8 $\pm$ 1.9   & 1.64 $\pm$ 0.23 &  0.7 \\  % D=222pc
\hline
\end{tabular}
\caption{Results of fitting the Mie scattering shell model with different dust species}
\label{tab:Mie}
\end{table*}

%5e35/((4/3.)*pi*(0.033*4.85e-6*195.*3e18)^3)
%\rouge{ ARTICLES PIERRE : la plume de CN est plus visible a 1.24  micron. 

\cite{2016A&A...585A..28K} published observations of Betelgeuse with the SPHERE/ZIMPOL instrument, finding an asymmetric envelope where the degree of linear polarization peaks at a radius of about three times the $K$-band limb-darkened radius, that is, $\approx$ 65 mas (see Fig.~7 of that paper).
Assuming the NIR dust shell that we characterized with SAMPol data is part of the same envelope seen in the visible, we can extrapolate the properties of our three dust models to visible wavelengths and compare them to the ZIMPOL measurements. 

\section{Comparison with visible polarimetric measurements}
\label{rad}

In order to construct a framework in which polarization models could be compared with both SAMPol and ZIMPOL data, the radiative transfer code MCFOST \citep{2006A&A...459..797P,2009A&A...498..967P} was employed. 
By constructing a set of wavelengths spanning a given filter bandpass, images were accumulated using weights given by the spectral transmission profile of the filter.
From the simulated images, model predictions could be derived for direct comparison with ZIMPOL and SAMPol observables.
%Stellar polarization was also included in the calculations following (publi du mail de C. Pinte).
%Mention the spectra? The spectrum we used is issued from convection simulation does not change anything. Andrea.

\begin{figure}[h!!]
\begin{center}
 \includegraphics[width=3.2in]{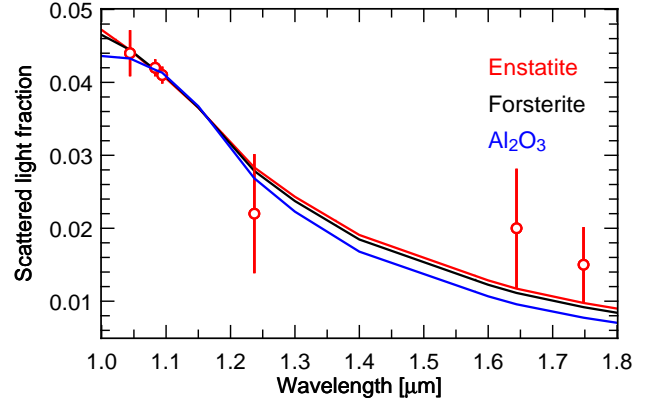}
\end{center}
\caption{Same as Fig.~\ref{scat_all_3}, but the curves for the scattered-light fraction were computed with the radiative transfer code MCFOST using the parameters listed in Table~\ref{tab:mcfost_param} and a Mie scattering. This accurately reproduces the parametric curves presented in Fig.~\ref{scat_all_3}.}
\label{gas2dust}
\end{figure}

As a sanity check, we ensured that we were able to reproduce the parametric modeling of the scattered-light fraction presented above. This is demonstrated in Fig.~\ref{gas2dust}, where the curves were produced using model parameters found in Table~\ref{tab:mcfost_param} and assuming Mie scattering in a thin (0.05 au wide) dust shell. The only free parameter was the dust shell mass, which we tuned to match the parametric modeling curves (Fig.~\ref{scat_all_3}). For dust composed of forsterite and alumina, these values were within the 1$\sigma$ uncertainty found on the dust shell mass with the parametric modeling (Table~\ref{tab:Mie}), and marginally worse for the enstatite. These new values for the dust shell mass are presented in Table~\ref{tab:mcfost_param}.

\begin{table*}
\centering
\begin{tabular}{lccl}
\hline
MCFOST Parameters &  values \\
\hline
Distance & 222 pc \\
Effective temperature & 3600K \\
Stellar radius & 900 R$_\odot$ \\
Stellar mass & 15 M$_\odot$  \\ 
\hline
Dust shell radius & 7.3 au \\ % 33 mas at 222pc
%Dust mass & 1.06e-10 M$_\odot$ (Alumina) ; 1.58e-10 M$_\odot$ (enstatite) ; 1.31e-10 M$_\odot$ (forsterite)  \\
Dust shell mass & 1.38e-10 M$_\odot$ (alumina) ; 2.02e-10 M$_\odot$ (enstatite) ; 1.72e-10 M$_\odot$ (forsterite)  \\
% using dust shell radius to match the parametric curves
%Dust shell mass & 1.49e-10 M$_\odot$ (alumina) ; 1.64e-10 M$_\odot$ (enstatite) ; 1.58e-10 M$_\odot$ (forsterite)  \\
Minimum grain size &  0.284 $\mu m$ (alumina) ; 0.324 $\mu m$ (enstatite) ;  0.310 $\mu m$ (forsterite)    \\ 
Maximum grain size &  0.298 $\mu m$ (alumina) ;  0.348 $\mu m$ (enstatite) ;   0.330 $\mu m$ (forsterite)\\  
 
%NB 3.74 & \\  // FILtre bizarrem barres d;erreur gigantesques
\hline
\end{tabular}
\caption{Parameters describing the star and dust shell models used for MCFOST simulations.} %$^1$: \cite{2008AJ....135.1430H}.  $^2$: }
\label{tab:mcfost_param}
\end{table*}    

\begin{figure}[h!!]
\begin{center}
 \includegraphics[width=3.2in]{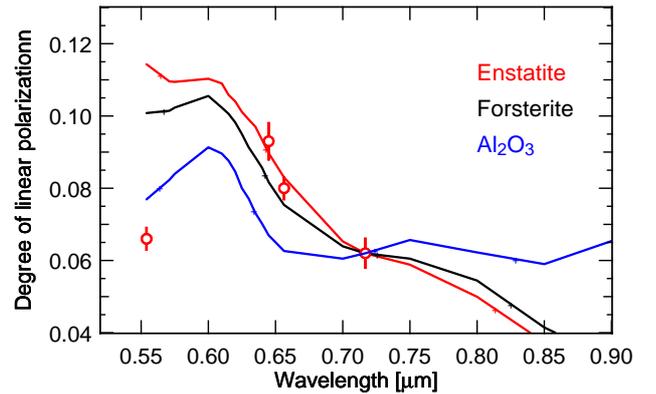}
\end{center}
\caption{Degree of linear polarization as a function of wavelength. Red circles indicate the maximum ZIMPOL data points reported in Fig.~7 of \cite{2016A&A...585A..28K}. With the goal of comparing their relative variation, the three dust models were offset with a constant so that their 0.717 $\mu m$ value matches the ZIMPOL datapoint (0.0012, 0.0077, and 0.0231 for enstatite, forsterite, and alumina, respectively). }
\label{degpol}
\end{figure}

Employing the same parameters we used previously to produce the Mie scattering parametric models, we computed the degree of linear polarization as described in \cite{2016A&A...585A..28K} in the four spectral filters.
%As can be seen in Fig.~\ref{degpol}, the three dust species behave more distinctively than the scattered light fraction in the NIR. Alumina and enstatite produce the lowest and highest values of degree of linear polarization, respectively. 
It is important to note that the maximum levels of the ZIMPOL degree of linear polarization are found at a radial distance between 60 and 80 mas. Therefore we cannot directly compare the absolute values with those predicted by our dust models, where the shell is located at around 33 mas. Absolute values for the degree of linear polarization are indeed very sensitive to the grain density, the dust shell radius, and thickness. However, we can compare the variation in functional form with wavelength because this is sensitive to the dust composition. From this plot, given in Figure~\ref{degpol}, it is apparent that the variation in the ZIMPOL maximum degree of linear polarization is better reproduced across the red spectral region by enstatite or forsterite, but the $V$-band data point is better reproduced by the alumina. 

%Can the spectral variation of deg lin pol be due to the brightness distribution of Betelgeuse, SED of Betelgeuse should peak around 0.83 microns --> do not explain the  variation since the peak is not there.]

MCFOST computes the dust temperature using the Monte Carlo method.
At 7.3 au, the dust grain temperatures are 1650 K, 1250K, and 750 K for the enstatite, forsterite, and alumina species, respectively.  For enstatite, the temperature seems too high by a few hundred Kelvin to allow for dust condensation \citep{2003A&A...408..193J}, even if the required constituent atoms were available in sufficient numbers and for a time that is long enough for grain growth. Taken at face value, this would therefore rule out an in situ formation for this dust shell located at 1.5 $R_{\star}$. The case is less clear for forsterite, but on the other hand, the condensation of alumina would seem definitely plausible based on the temperature
value alone.

Given its extreme simplicity, the model shows promise that better matches might be obtained with the addition of one or a combination of further factors. We list these factors below.
\begin{itemize}
\item{There may be a dust species or a combination of dust species that we did not consider here and that better reproduces the ZIMPOL measurements.}
\item{Some effects that we did not model with MCFOST (non-spherical grains, hollow spherical grains, etc) could alter the outcomes.}
{\item{The $V$-band filter has a larger spectral bandwidth (80 nm) than the other filters. Several chromatic effects, both instrumental and/or astrophysical, may be at play to generate a lower degree of polarization for this datum.}}
\item{The structure imaged by ZIMPOL may not have the same composition and characteristics as the dust seen in the NIR shell. The ZIMPOL polarized images show that the dust atmosphere is clearly extended and deviates from the thin dust layer that we used to compare with the NIR measurements.}
\end{itemize}

%We have not tried a combination of them....

\section{Discussion}
\label{discuss}
\subsection{Polarization from the stellar atmosphere}
The stellar atmosphere is another potential source of polarization. It is therefore important to investigate its effects on the observables we analyzed previously.

Center-to-limb polarization variations (CLPVs) of stellar atmosphere calculations in F, G, and K type stars were presented in \cite{2016A&A...586A..87K}.
We adapted these calculations to the case of Betelgeuse and used the following stellar parameters: M = 15 M$_{\odot}$, R = 1000 R$_{\odot}$, T$_{eff}$ = 3600K, and log g=0.0.

\begin{figure}[h!!]
\begin{center}
 \includegraphics[width=3.8in]{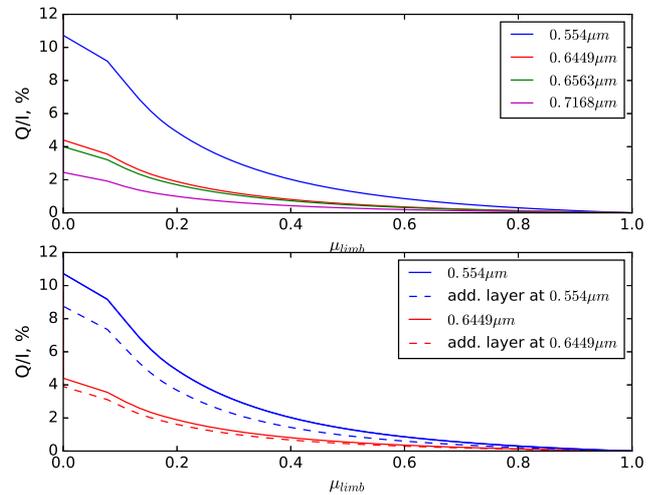}
\end{center}
\caption{Limb-darkening polarization, intensity-normalized Stokes Q parameter as a function of $\mu_{limb}$. Upper panel: These curves are presented in four filters for the stellar atmosphere only. Lower panel: An additional molecular layer has been added to the stellar atmosphere for two of the filters.}
\label{degpol_comp}
\end{figure}

The upper panel of Figure~\ref{degpol_comp} shows the limb-darkening fractional polarization intrinsic to the stellar atmosphere as a function of $\mu_{limb}$ , which corresponds to $\mu$  ($\mu=cos(\theta),$ where $\theta$ is the angle between the line-of-sight direction and the direction normal to the stellar atmosphere), except that it ranges from where the largest gradient of the center-to-limb variation of intensity is observed \citep[definition of the position of the observed stellar limb, see section 3.1.2 of][]{2016A&A...586A..87K}. The position of the observed stellar limb then corresponds to $\mu_{limb} = 0$ and the normal to the atmosphere corresponds to $\mu_{limb} = 1$. This plot shows that the limb-darkening fractional polarization has a declining strength to longer wavelengths (plots are tailored for the specific ZIMPOL filters we employed).

The lower panel reveals the effect on this quantity of a molecular layer (optical thickness of 0.01, distance of 1.5 stellar radius) in addition to the stellar atmosphere for two selected filters: multiple scattering causes the fractional polarization to decrease up to a few percent depending on the wavelength.

\begin{figure}[h!!]
\begin{center}
 \includegraphics[width=3.5in]{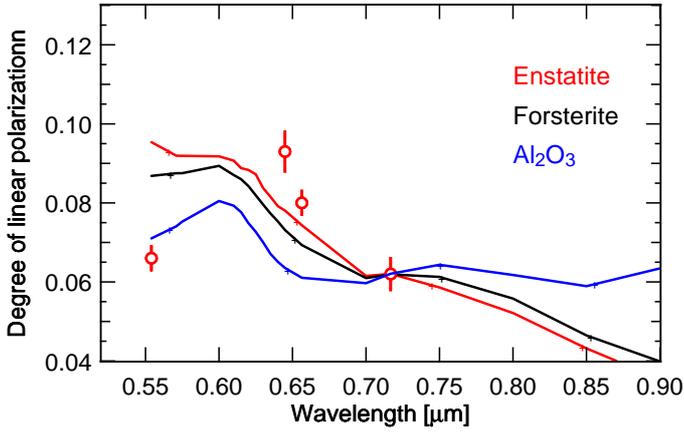}
\end{center}
\caption{Same as Figure~\ref{degpol}, with the addition of polarization stemming from the stellar atmosphere.}
\label{degpol_LD}
\end{figure}

We next integrated these CLPVs into our MCFOST models to take both the stellar atmosphere and the dust shell in the net polarization computation into account.
Figure~\ref{degpol_LD} shows the same quantity as Fig.~\ref{degpol}, but the CLPV is now included in the calculation of the Stokes parameters.

The main consequence of the addition of the CLPVs into our models is a global decrease in the degree of linear polarization at the visible wavelengths.
As demonstrated in the lower panel of Fig.~\ref{degpol_comp} with a molecular layer, this decrease can be attributed to the multiple scattering that takes place in a layer that is added to the stellar atmosphere. All models now clearly underestimate the observation data points at 0.6449 $\mu m$ and 0.65634 $\mu m$.  However, with this global decrease, the alumina model is able to better reproduce the 0.554 $\mu m$ data point. Several explanations for this inclusion of the CLPVs can be invoked to explain the overall poorer modeling of the ZIMPOL degree of linear polarization:

\begin{itemize}
\item{The degree of linear polarization observed with ZIMPOL contains a very weak direct polarization signal from the stellar atmosphere because the wider circumstellar dust is well separated from the star at the instrumental resolution.}
\item{The models with the lowest value of log g have log g=0.0, while Betelgeuse has log g in the [-0.2: -0.5] range. }
\item{CLPV calculations were made for a homogeneous atmosphere, which (at some level) is not the case for Betelgeuse. A modeling effort taking the temperature distribution predicted by hydro-radiative simulations into account is ongoing.}
\end{itemize}
We conclude that the additional complexity of adding stellar photosphere CLPVs to the modeling has done little to improve the ability to reproduce the wavelength dependence that is evident in the ZIMPOL data.

% A mettre ou pas?
\subsection{Shell extension and mass-loss rates}
The dust shell masses reported here are lower than the mass-loss rates derived from infrared excesses \cite[e.g.,][]{2005A&A...442..597V}.
% ###### It would be useful here to actually quote the literature mass loss rate. Just how much smaller is our shell??
%and little compare to extreme dust RSG producer like VY CMa \citep[> 1e-4][]{2015ASPC..499..335O}.
However, interferometric polarimetry is most sensitive to thin dusty structures in close proximity to the star because they will provide the 
strongest polarized signal. 
The thin dust shell model invoked here might even represent a dense inner rim of a more extended envelope whose full extent is not easily revealed with this technique. This would explain why we see a thin dust shell and not a continuous structure as found by ZIMPOL, and also why we derive a relatively low mass-loss rate. We may only witness a part of the dust population located at the inner base of a larger envelope.

\begin{figure}[h!!]
\begin{center}
 \includegraphics[width=3.7in]{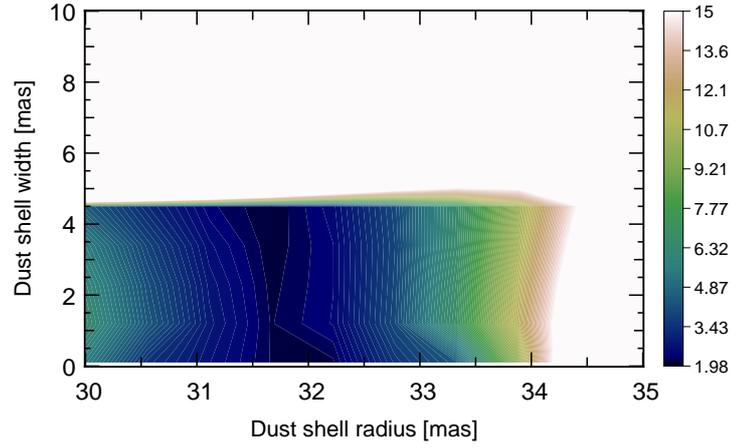}
\end{center}
\caption{Map of reduced $\chi$-squared for an extended dust shell model at 1.09 microns.}
\label{chi2map_ext}
\end{figure}

We attempted to model the Vh/Vv data (see section 3.1) using an extended shell model with a Gaussian density distribution. The reduced $\chi$-squared map is presented in Fig.~\ref{chi2map_ext}.
The best $\chi$-squared values we achieved are higher than for the thin shell model, even though we added the thickness (Gaussian FWHM) parameter. We note a large drop in $\chi$-squared values above a shell thickness of $\sim$4.5 mas ( $\sim$1 au at 222 pc). For this extended model, the best fit to the data is obtained for apparent dust shell radii lower than the values reported in Table~\ref{tab:dustshell}.  We find that an apparent dust shell radius of 32 mas and a shell thickness of 4.5 mas with an FWHM lead to a density decrease of $\sim$10\% in order to reproduce SAMPOL data, whereas the size of the dust grain has to be increased by 20\% at most.

\subsection{Comparison with AGBs}
The many similarities of the astrophysical environments make a comparison between AGBs and RSGs very relevant in interpreting the properties of dust found close to the stellar surface.
Recent studies, particularly using visible and NIR polarimetry, have shown that dust with  grain sizes of a few tenths of a micron is common below 2 $R_{\star}$; this conclusion  seems to hold for varying mass-loss rates \citep{2012Natur.484..220N,2015A&A...584L..10S,2016A&A...589A..91O, 2016A&A...591A..70K}. 
We here add similar findings for the case of Betelgeuse. A common mechanism to explain dust nucleation and formation, despite differences in temperature, chemical composition, and stellar dynamics, might be indicated. 
Multi-epoch campaigns now offer the opportunity to link grain size variations with the AGB pulsation cycle \citep{2017A&A...597A..20O}.
However, in the case of RSGs, no strong pulsation analogous to that in AGBs is present, and time variability is rather dominated by convection time scales.
A further extension of our study would therefore be to quantify the time variability of dust characteristics in the first stellar radii of Betelgeuse.

% A FAIRE : clotures de phase en fonction du PA
%The importance of Rayleigh scattering on molecules has a dependency in $\lambda^{-4}$. discussed in http://adsabs.harvard.edu/abs/1973MNRAS.163..261T
%\rouge{Is 1G enough to align dust grain and make the polarization to increase ?}
%\rouge{how about magnetic field , could it directly create a polarization ?}
%Checking how 3d convection/hydro models of red supergiant atmospheres could allow the conditions to form dust is left for a future work.
%polarized signal could be interstellar. But the reported values in the literature can't account for a signal high as 5\% \rouge{TBC}.

    \section{Conclusion}
    \label{conclusect}

We reported on interferometric measurements of \object{Betelgeuse} that were obtained with NACO/SAMPol.  We detected the polarized signature interpreted as a scattered-light shell located at $1.5~R{\star}$ using a thin dust shell model. A model that includes a $\sim$4.5 mas extension of the shell would result in a 10-20\% variation in the dust density and grain size.
Although the data constrain the shell and size of its constituent grains, they cannot alone disentangle the composition of dust. However, such a study can be attempted using comparison of the dust shell models with SPHERE/ZIMPOL measurements. We tend to favor the iron-free silicates such as forsterite for the dust composition rather than alumina, although no single species was able to perfectly reproduce the data. It is possible that the new dusty structure reported here could form the base of a scattering-driven wind following the theoretical predictions reported in \cite{2008A&A...491L...1H}.  In order to proceed in the characterization of the inner dust shell of \object{Betelgeuse}, polarimetric observables should be monitored in more spectral bands, particularly in the visible continuum. Close to the photosphere, it is likely that the dust shell is affected by changes in surface luminosity that are due to convection and so variability at timescales of some months is expected.

%\rouge{TBD}
%Taking into account the geometry and orientation of the dust could help understand the formation of this density structure. Interferometric image reconstruction of density structures of SAMPol is left for a future publication. 
 %Modelling including grain growth and dynamics

   \appendix
 %  \section{Appendix}

\renewcommand{\thefigure}{A.\arabic{figure}}
\setcounter{figure}{0}

\begin{figure*}[h!!]
  \includegraphics[width=3.5in]{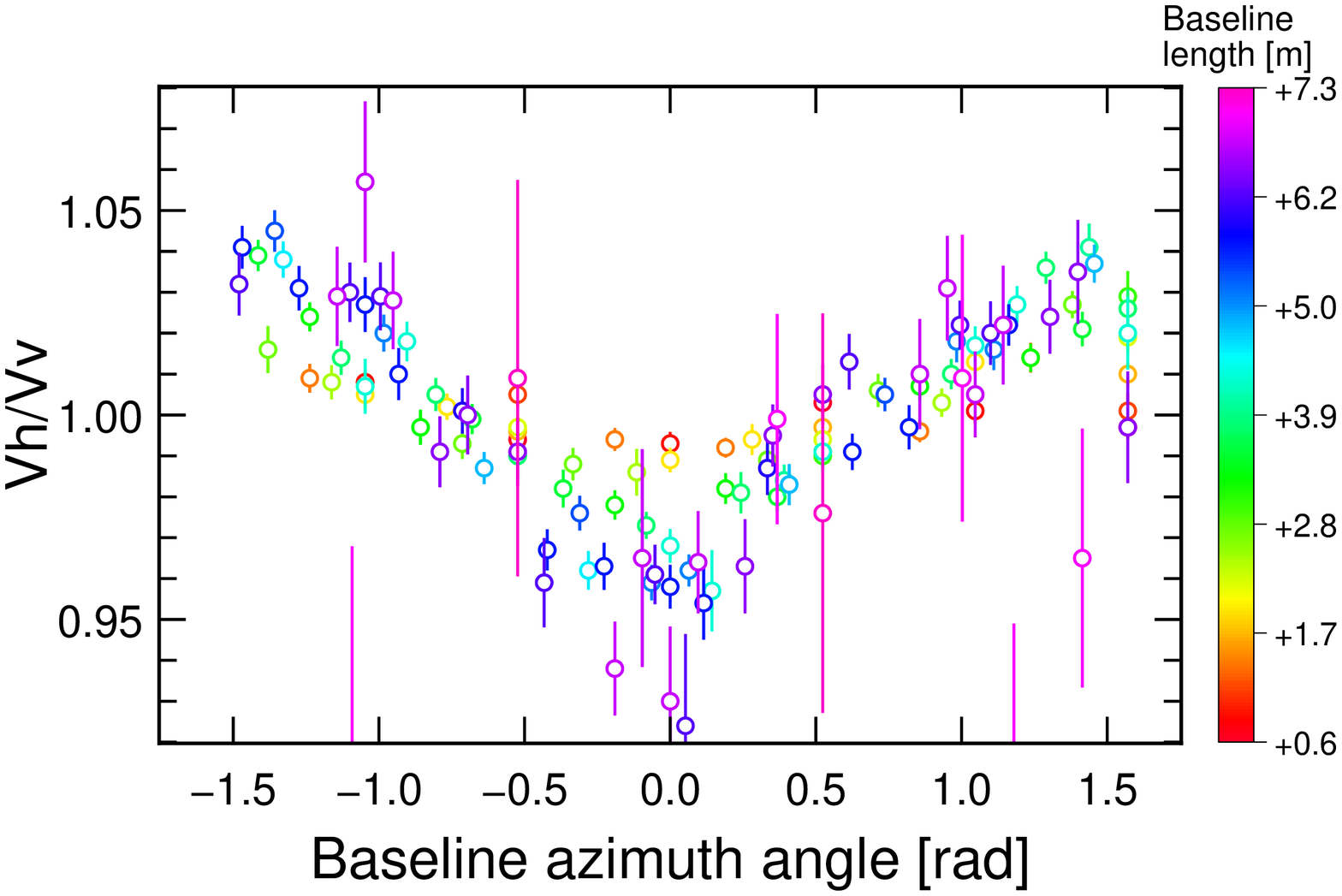}
\begin{center}
 \vspace{3mm}
    \includegraphics[width=3.5in]{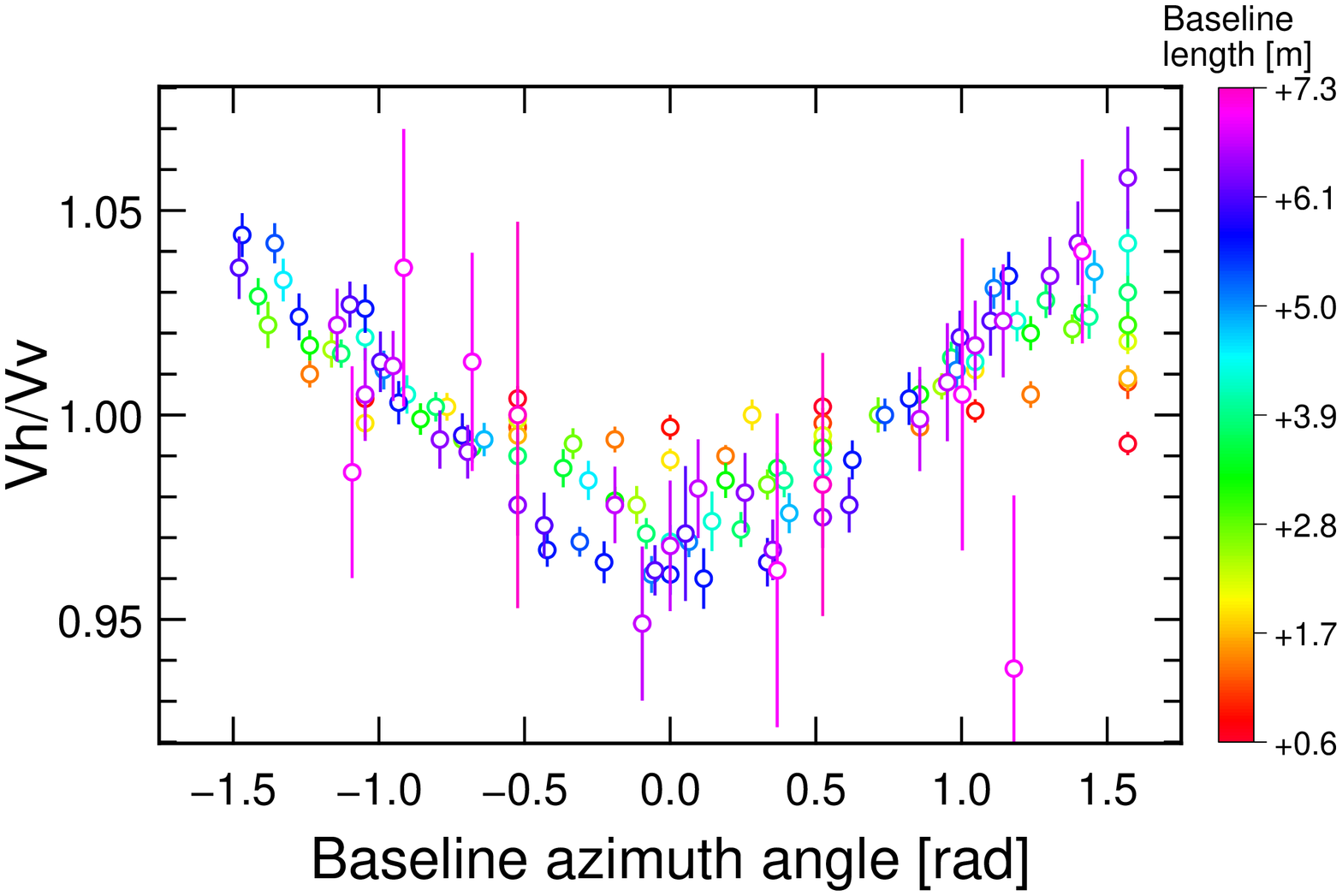}
 \hspace{3mm}
    \includegraphics[width=3.5in]{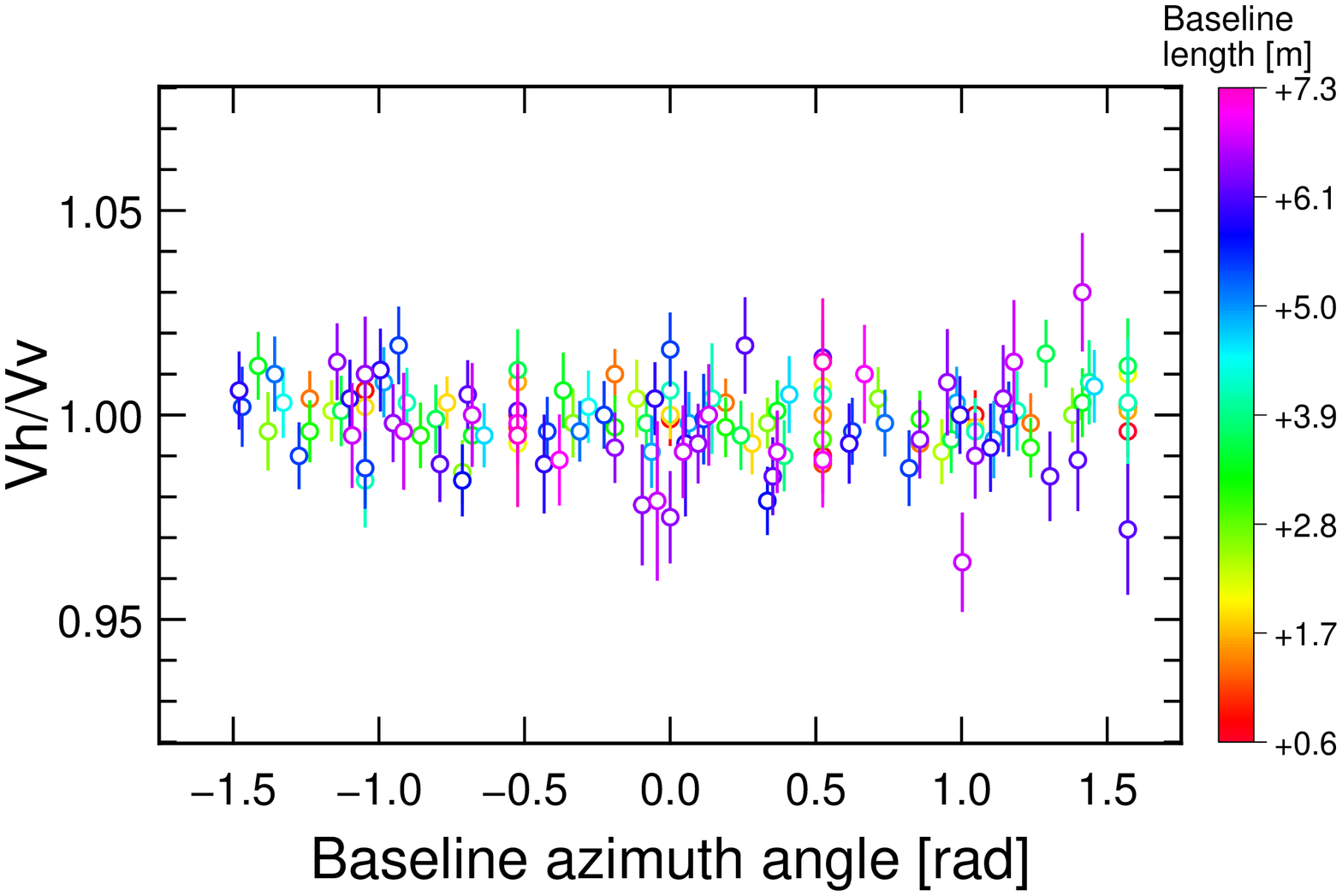}
    \hspace{3mm}
 \includegraphics[width=3.5in]{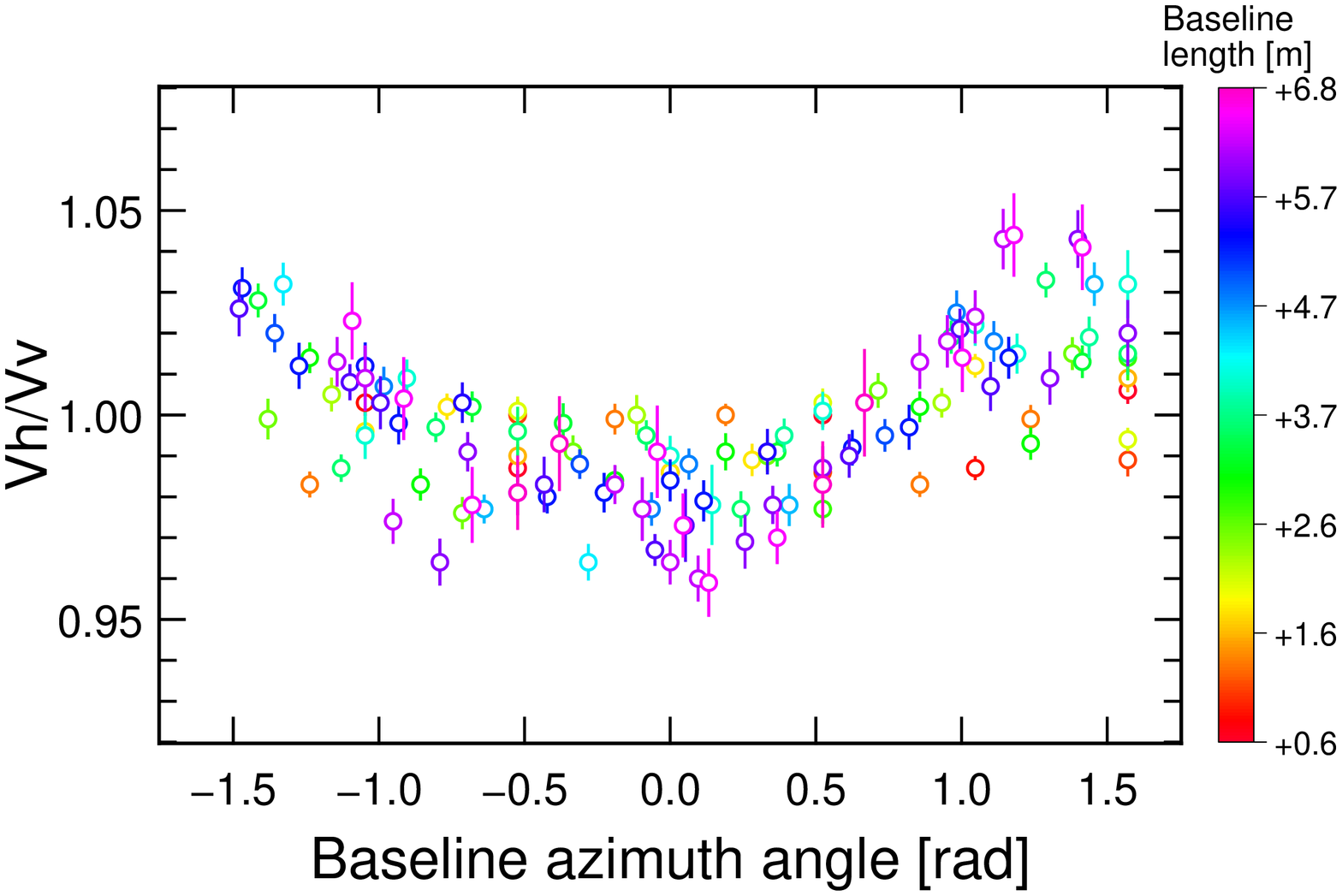}
 \hspace{3mm}
 \includegraphics[width=3.5in]{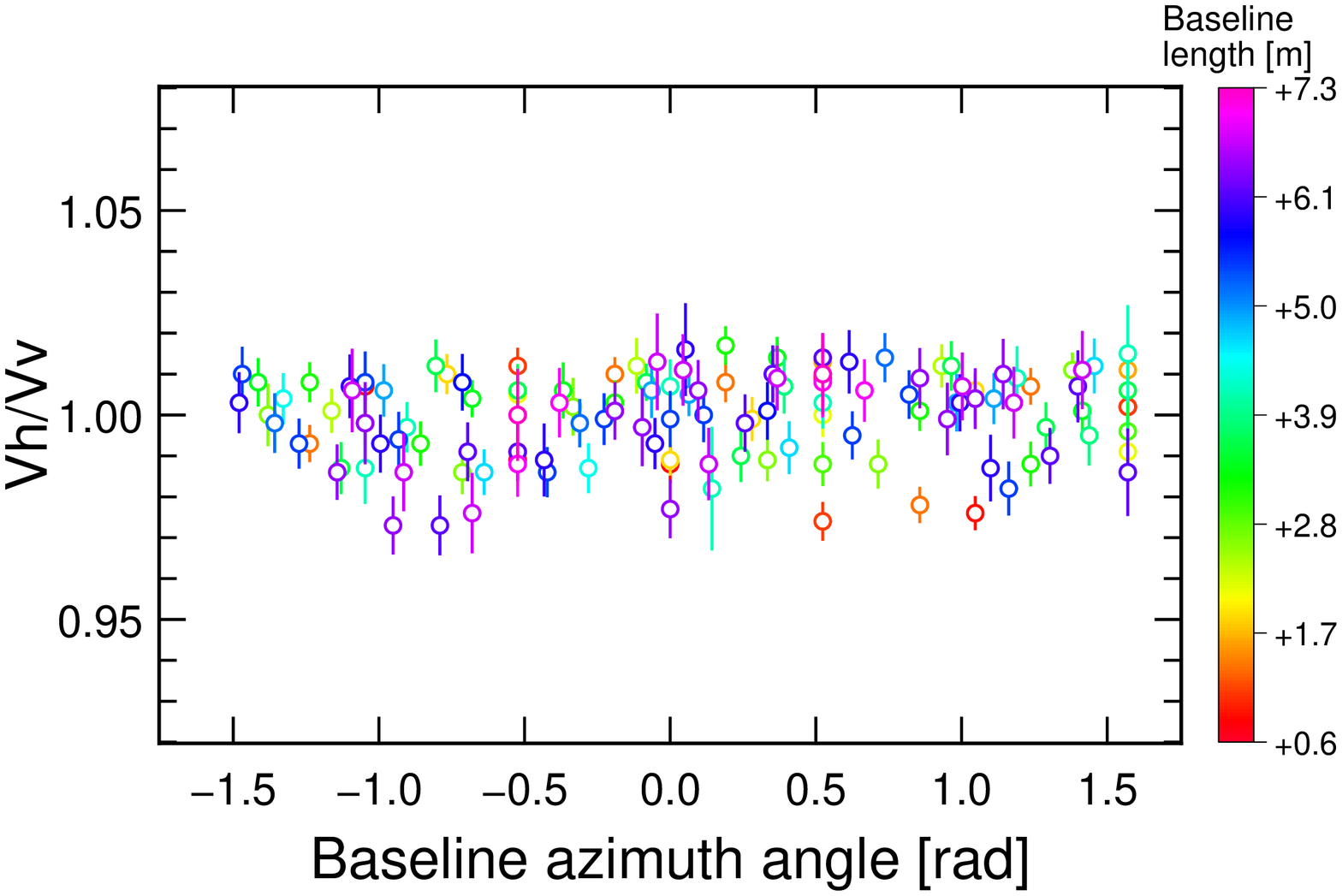}
 \end{center}
\caption{Differential visibility ratio plotted as a function of azimuth angle, color-coded with baseline length. Left: Betelgeuse. Right: Aldebaran. From top to bottom: Filters centered on 1.08, 1.09, and 1.24 $\mu$m. We did not observe Aldebaran in the 1.08 filter.}
\label{vhvv1}
\end{figure*}

\begin{figure*}[h!!]
\begin{center}
  \includegraphics[width=3.5in]{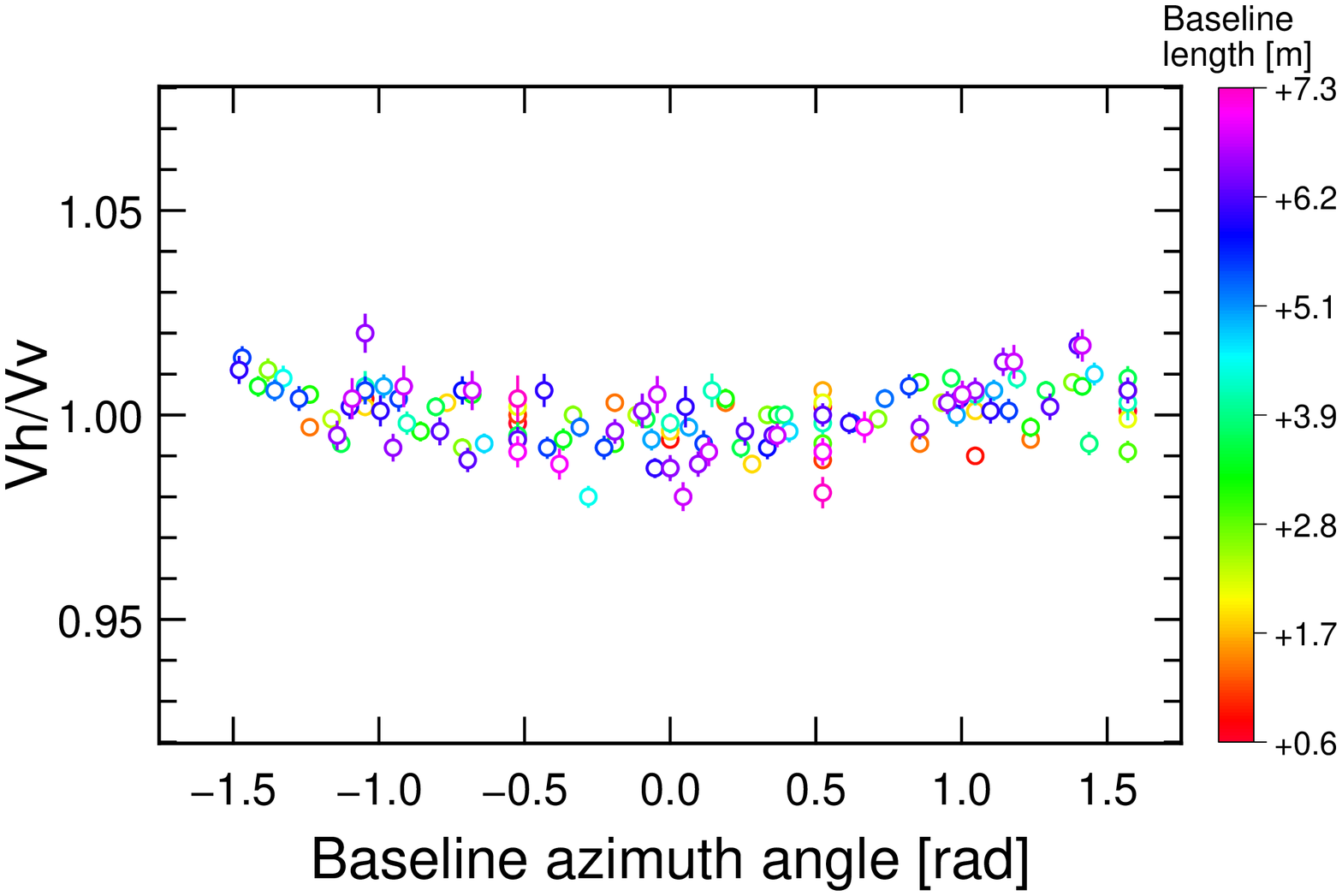}
   \hspace{3mm}
   \includegraphics[width=3.5in]{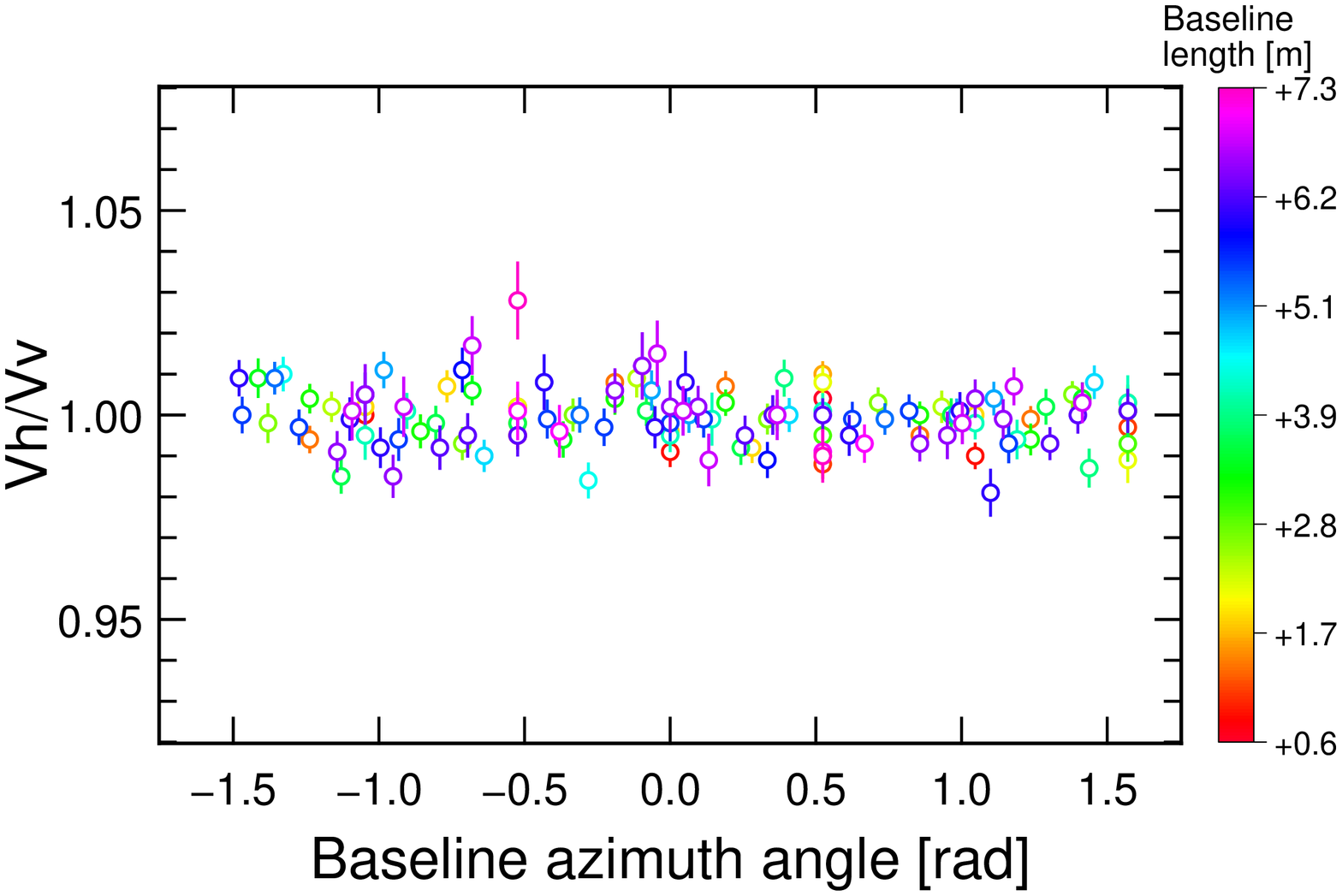}
 \vspace{7mm}
    \includegraphics[width=3.5in]{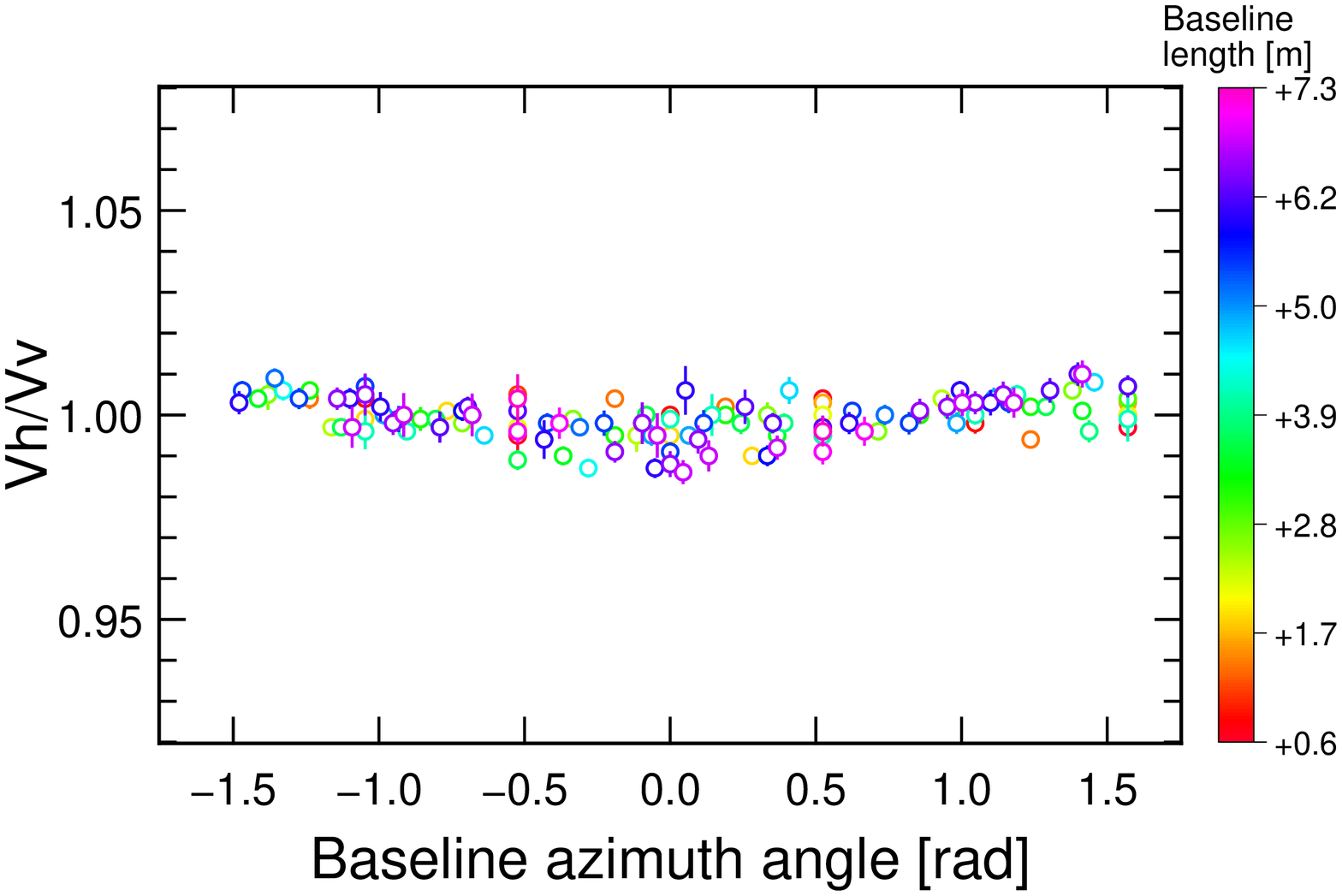}
 \hspace{3mm}
    \includegraphics[width=3.5in]{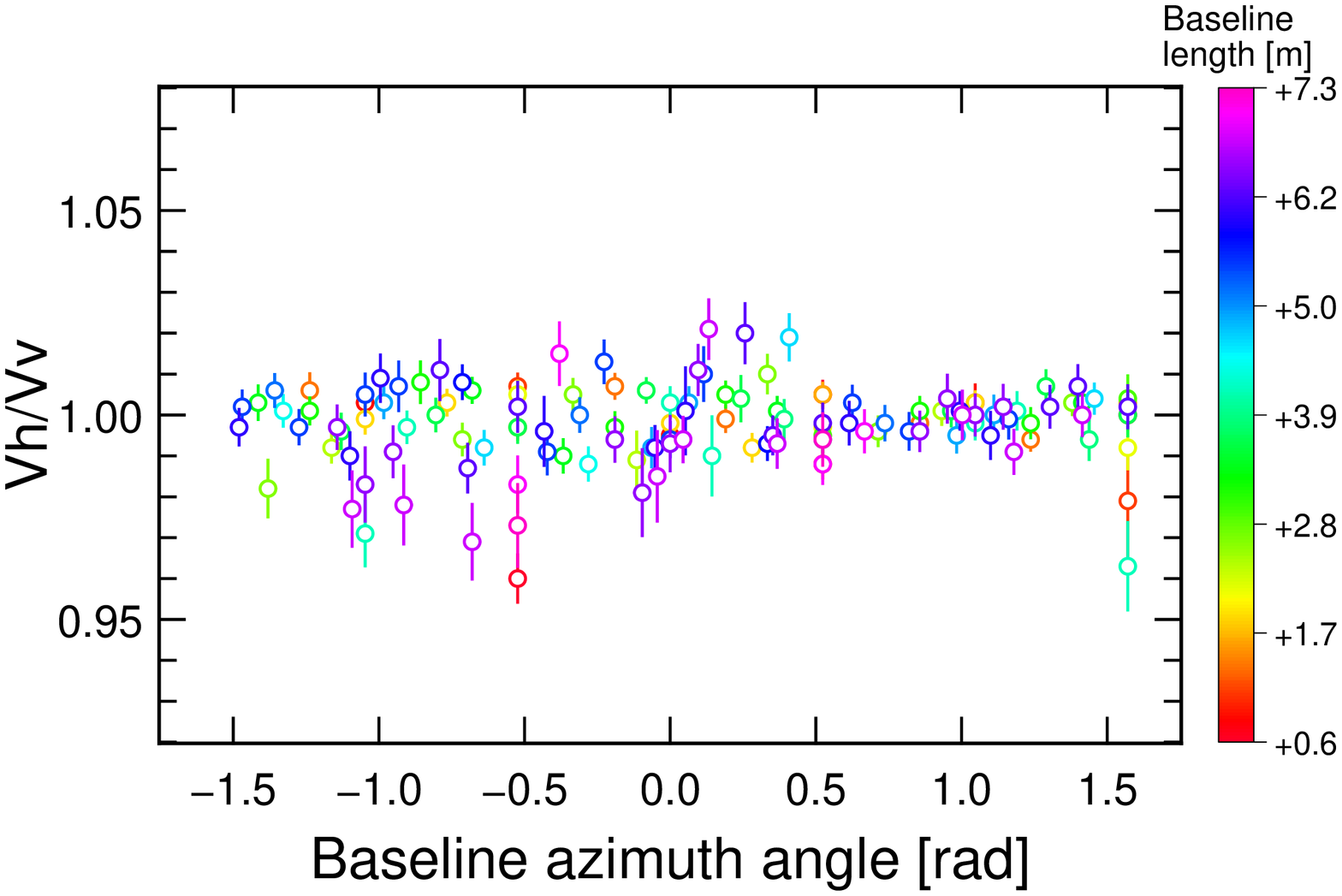}       
\end{center}
\caption{Same as Fig.~\ref{vhvv1} for filters centered on 1.64 (top) and 1.75  $\mu$m (bottom).}
\label{vhvv2}
\end{figure*}

   \begin{acknowledgements}
This research has made use of the Jean-Marie Mariotti Center \texttt{SearchCal} service \footnote{Available at http://www.jmmc.fr/searchcal} that is co-developed by FIZEAU and  LAOG/IPAG, and of CDS Astronomical Databases SIMBAD and VIZIER \footnote{Available at http://cdsweb.u-strasbg.fr/}. XH thanks Cornelis Dullemond for the time he spent supporting preliminary simulations on RADMC3D as well as Susanne H{\"o}fner and  Bernd Freytag for helpful discussions.
   \end{acknowledgements}

\end{document}